\documentclass[10pt,conference]{IEEEtran} 
\IEEEoverridecommandlockouts
\usepackage{cite}
\usepackage{amsmath,amssymb,amsfonts}
\usepackage{graphicx}
\usepackage{textcomp}
\usepackage{xcolor}
\def\BibTeX{{\rm B\kern-.05em{\sc i\kern-.025em b}\kern-.08em
    T\kern-.1667em\lower.7ex\hbox{E}\kern-.125emX}}
    
\newcommand{\TODO}[1]{\textcolor{red}{#1}\GenericWarning{}{LaTeX Warning: TODO: #1}}\newcommand\todo\TODO
\newcommand{\etal}{\textit{et\,al.}\xspace}
\newcommand{\ie}{\textit{i.e.},\xspace}
\newcommand{\eg}{\textit{e.g.},\xspace}
\usepackage{xspace}

\usepackage{hyperref}

\usepackage{algorithm}
\usepackage{algorithmicx}
\usepackage{algpseudocode}

\usepackage[squaren,Gray]{SIunits}    

\begin{document}
\title{Can~We Spot Energy~Regressions using Developers~Tests?}

\author{
\IEEEauthorblockN{Benjamin Danglot}
\IEEEauthorblockA{\textit{Research \& Development} \\
\textit{Davidson Consulting}\\
Paris, France\\
benjamin.danglot@davidson.fr}
\and
\IEEEauthorblockN{Jean-Rémy Falleri}
\IEEEauthorblockA{\textit{Univ. Bordeaux, Bordeaux INP, CNRS} \\
\textit{LaBRI, UMR 5800, F-33400 Talence, France} \\
\textit{Institut Universitaire de France} \\
falleri@labri.fr}
\and
\IEEEauthorblockN{Romain Rouvoy}
\IEEEauthorblockA{\textit{Univ.\,Lille / Inria / IUF}\\
Lille, France \\
romain.rouvoy@univ-lille.fr}
}

\maketitle

\begin{abstract}
\paragraph*{Background}
\emph{Software Energy Consumption} (SEC) is gaining more and more attention. 
In this paper, we tackle the problem of hinting developers about the SEC of their programs in the context of software developments based on \emph{Continuous Integration} (CI).
\paragraph*{Objective}
In this study, we investigate if the CI can leverage developers' tests to perform a new class of test: the energy regression testing.
Energy regression is similar to performance regression, but focused on the energy consumption of the program instead of standard performance indicators, like execution time or memory consumption.
\paragraph*{Method}
We propose to perform an exploratory study of the usage of developers' tests for energy regression testing.
We propose to first investigate if developers' tests can be used to obtain stable SEC indicators.
Then, to consider if comparing the SEC of developers' tests between two versions can accurately spot energy regressions introduced by automated program mutations.
Finally, to assess if it can successfully pinpoint the source code lines guilty of energy regressions.
\paragraph*{Impact}
Our study will pave the way for automated SEC regression tools that can be readily deployed inside an existing CI infrastructure to raise awareness of SEC issues among practitioners.
\end{abstract}


\section{Introduction}
\label{sec:introduction}
\emph{Software Energy Consumption} (SEC) is gaining more and more attention from both researchers and practitioners.\footnote{\url{https://greensoftware.foundation}}
However, the awareness of SEC among developers is not yet widely spread and is just starting to be considered as a key indicator of a project quality.

In this paper, we tackle the problem of hinting developers about the SEC of their programs.
More specifically, this study takes place in the context of Agile development, DevOps, and \emph{Continuous Integration} (CI).
In particular, we aim to flag commits that negatively impact the SEC of an application under development or maintenance.

Whenever a developer pushes changes as a commit, the CI toolchain is in charge of building and checking if the submitted changes are "correct".
Traditionally, CI performs regression testing---\ie it verifies the absence of regression bug~\cite{beller2017oops}.
A regression bug is an unintended loss of features introduced by code changes, which were not intended to modify this part of the program behavior.
As code changes may have side effects that introduce an involuntary and undesired change of the behavior (or bug), these regression bugs can be spotted by the tests triggered by the CI.
The most simple regression detection techniques is to execute all the tests, no matter the submitted changes.
If any test fails, this means that there is a regression and the associated code changes are then tagged as ``breaking''.
Then, the developer who authored the code changes must fix the regression bug(s) and make all the tests to pass again.

In this study, we investigate if the developers' tests~\cite{meszaros2007xunit} can be leveraged to perform an additional verification as part of the CI: the energy regression testing.
Our idea is inspired by the work of Ding~\etal~\cite{ding2020towards}, who showed that such tests can already spot performance regressions in a program.
Energy regression is similar to performance regression, but focused on the energy consumption of the program instead of its execution time or memory consumption.
In other words, we aim to integrate in the CI an energy regression oracle that assess that the applied code changes do not severely degrade the SEC.
In particular, the CI would tag a code change as ``breaking'' if it severely increases the energy consumption of the program.
In addition to this, we envision that the approach would be able to list candidate atomic code changes (\eg lines of code that have been changed) that might cause the observed SEC drift.

Similarly to performance, energy consumption is a dynamic property: one must execute the program to collect SEC indicators.
As we aim to detect the negative impact on the energy consumption, we need to compare the energy consumption of two versions of a program: the version before applying the commit and the one after applying the commit.
To do so, developers' tests can be considered as a candidate workload, as they are already executed for each version, and can be used to monitor dynamic properties' variations, as long as these tests are not modified by the commit.

In this paper, we take a first step towards this vision by performing an exploratory study of the applicability of such an approach.
We first investigate if stable SEC indicators can be computed from developers' tests.
Second, we investigate if comparing SEC indicators of two version can accurately spot energy regressions introduced by automated program mutations. 
We also study the difference of SEC indicators along the commit history of several real-world projects.
Finally, we investigate if the computation of SEC indicators, used together with fault localization techniques, can successfully pinpoint the guilty code lines, using program mutations as a ground truth.

\section{Research Questions}\label{sec:research_questions}
Our goal is to capture the impact of code changes on the energy consumption of any software.
To achieve this, we need to compare the energy consumption of two versions of the same program: the version before applying the code changes and the version after applying the code changes.
To compare these two versions, we can compute the ``energy delta'' ($\Delta_E$), which is the difference of energy consumptions: $\Delta_E = E(v_2) - E(v_1)$.
This $\Delta_E$ will indicate if the code changes increase the energy consumption of the program.
In addition to this, we aim at providing hints to the developer on the code changes (\eg a specific changed line of code) that is potentially causing the energy consumption increase.
In the remaining of this paper, these versions of the program will be referred as $v_1$ and $v_2$ for the version before applying the code changes and the one after applying the code changes, respectively.

The goal of this study is to explore if we can exploit existing developers' tests to qualify code changes according to their impact of the software energy consumption.
In this context, qualify means to indicate if they increase or decrease the energy consumption, and to quantify this variation regarding to global software energy consumption.
Our goal is to warn the developers whenever the introduced code changes are severely increasing the energy consumption of the program.
In other words, we propose a new test infrastructure that supports developers by detecting energy regressions within the continuous integration and pinpointing the suspicious code.

\newcommand{\RQOne}{Can the energy consumption of tests be measured in a stable fashion?}
\newcommand{\RQTwo}{Can the tests be used to a detect potential energy regression introduced by code changes?}
\newcommand{\RQThree}{Can the tests be used to highlight atomic code changes that are causing an energy regression?}

To sum up, we aim at answering to the following research questions:
\begin{itemize}
    \item [RQ$_1$] \RQOne{}
    \begin{itemize}
        \item [H$_1$] Measuring the energy consumption of tests produces stable measurements.
        \item [H$_2$] Measuring dynamic performance indicators, such as duration or executed instructions of tests, can be used to approximate the energy consumption and produces stable measurements.
    \end{itemize}
    This research question aims at verifying that tests can be exploited to measure the energy consumption of a small portion of code in a stable manner.
    It will deliver recommendations on which measurements can be taken in consideration to do so, by studying their stability across several measurements.
    \item [RQ$_2$] \RQTwo
    \begin{itemize}
        \item [H$_3$] A potential energy regression introduced by code changes can be detected from the tests by computing the energy difference of unmodified tests executed before and after the code changes.
    \end{itemize}
    \item [RQ$_3$] \RQThree
    \begin{itemize}
        \item [H$_4$] The tests can help to locate code changes that could cause an energy regression, based on the lines executed by the tests and their energy consumption.
    \end{itemize}
\end{itemize}

\section{Variables and Experimental Protocol}\label{sec:variables}
In this study, we need to measure the energy consumption of a program for a given version.
To do this, we propose to use the following variables:
\begin{itemize}
    \item The energy consumed by the execution of a test on a given version of the target program, expressed in \emph{joules}~(J);
    \item The duration of the execution of a test on a given version of the targeted program, expressed in \emph{seconds}~(s);
    \item Several performance counters monitored along the execution of a test on a given version of the targeted program: 
        \textsf{instructions},
        \textsf{cycles},
        \textsf{cache-references},
        \textsf{cache-misses},
        \textsf{branches},
        \textsf{branch-misses}.
        These variables are absolute numbers.
\end{itemize}
The energy consumed by the execution of a test is the key metrics of our study.
The execution duration of a test is another candidate metrics to capture the consumption of a test. 
We believe that other counters, such as the number of instructions, cycles, etc. are also relevant candidate metrics to approximate the consumption of a test.
We argue that coupling these indicators to the energy consumption, expressed in joules, would allow us to consolidate the result and assess their stability.
These variables will be used to approximate the energy variations of a program from the tests.

In addition to these variables, we highlight various other variables related to our study:
\begin{itemize}
    \item The subset of tests that execute the lines of code changed by the commit.
        This variable is a list of tests;
    \item The modified line coverage per test.
        This variable is the function $exec(l,t)$ that takes a modified line $l$ and a test $t$ and returns a number of times that the test $t$ executes the line $t$.
\end{itemize}
The subset of tests that execute the code changes is relevant for our study, as they will be used to compute the difference of energy consumption between two versions of the same program.

\subsection{Energy Delta}
Using these variables, we will compute the energy delta of consumption related to some code changes.
This delta will capture the differences in terms of energy consumed, time execution, and others counters between the two versions of the same program, for each test.
The delta will be computed as follow:
\begin{equation}
    \forall t \in TS, \Delta_V(t) = V(t_{v2}) - V(t_{v1})
\end{equation}
where $TS$ is the subset of the test suite that executes the code changes, $t_{v2}$ and $t_{v1}$ are the execution of the test $t$ for the program after the commit and before the commit, respectively, and $V()$ is the monitored value for the given variable $V$---\ie the energy consumption, the time execution or performance counters.

\subsection{Experimental Protocol}
\label{sec:experimental_protocol}
In this section, we expose the experimental protocol that we plan to follow to answer to our research questions.

\subsubsection{Answer to RQ$_1$: \RQOne}
To answer RQ$_1$, we plan to monitor the energy consumption and performance indicators of tests that will be selected automatically.
To consider these metrics, we need to execute several times the tests to assess their stability, as developers' tests are usually small and fast, which might be a challenge to report on a stable measurement of the energy consumption.
While our approach can exploit any kind of tests---unit, integration, or performance tests---we argue that taking unit tests as several benefits.
In particular, unit tests are much more common than performance tests, allowing to apply the approach on a wider set of software.
While performance tests are also valuable for our purpose, their availability is often limited and they use to require system-level workloads.
In order to assess the stability of our measurement, we will compute statistical indicators, such as standard deviations, coefficient of variation and quartile coefficient of dispersion.
These indicators will deliver more robust insights on the stability of the collected metrics, and thus validate or not the hypotheses H$_1$ and H$_2$.
Having several measurements of energy consumption and dynamic performance indicators will provide us a way to compute a confidence interval that will indicate if the metrics are stable enough.

\subsubsection{Answer to RQ$_2$: \RQTwo}
To answer RQ$_2$, we will consider open-source projects from the wild.
Then, we propose to apply controlled mutations to generate new versions of the programs under study consuming more energy.
Then, for each injected mutation, we measure the energy consumption of the tests executed on the version before the mutation ($v_1$) and the energy consumption of the same tests executed on the mutated version ($v_2$).
We then compute delta $\Delta_E = E(v_2) - E(v_1)$.
If $\Delta_E > 0$ for all the cases, we can adopt developers' tests to detect energy regressions introduced by code changes, thus validating our hypothesis H$_3$.
In addition to the injected mutations, we plan also to compute the $\Delta_E$ on the commit history---\ie the list of all commits---of the selected projects.
From this, we will compute the ratio of code changes that will be considered as ``breaking''---\ie the code changes that severely increase the energy consumption.
This is done to check that not all the code changes are considered as ``breaking''.

As goes for RQ$_1$, the energy consumption measurements can be unstable.
To overcome this, we plan to:
First, apply the same strategy that we apply for RQ$_1$, by considering several measurements to overcome potential instability and compute confidence intervals over several runs.
Second, observe the intersection between the two ranges of energy measurements for $v_1$ and $v_2$.
If the intersection is empty, it means that the difference is likely due to the code changes. 
If the intersection is not empty, the difference might come from fluctuation in our measurements and we cannot conclude on the impact of the commits  to the variation of the energy consumption.

\subsubsection{Answer to RQ$_3$: \RQThree}
To answer RQ$_3$, we will consider open-source software projects from the wild.
Then, for each commit, we will compute the SEC difference between the two versions of the program.
We discard the case where the computed $\Delta_E > 0$.
When $\Delta_E \leq 0$, we will mutate the code, as done in RQ$_2$, to increase the energy consumption associated to the code changes.
Then, we will use the tests, the executed modified lines and their energy consumption to rank the code changes according to their suspiciousness.
This analysis should prioritize the changes related to our mutations as $\Delta \leq 0$, therefore the manual code changes committed by the developers should not be suspected (or at least, less suspected than the code changes done by the mutation).
If this is the case, one can assume that our analysis correctly assigns the cause of energy regression to the injected mutations.
If not, the analysis fails to spot the correct code changes that are causing the energy drift.
In the former case, our hypothesis H$_4$ will be validated, in the latter case our hypothesis will be rejected.

\subsection{Statistical indicators}
As the energy consumption, and the variables used to approximate it, exhibit some uncertainty, we need to perform several measurements to converge towards stable values.
To assess the stability of the variables, we propose to compute 3 indicators:
\begin{enumerate}
    \item \emph{standard deviation} is a measure of the amount of variation of a set of values. A low standard deviation indicates that the values tend to be close to the mean of the set, while a high standard deviation indicates that the values are spread out over a wider range.
    The standard deviation $\sigma$ of a population of ($N$) is computed by the following equation:
    \begin{equation}
       \sigma = \sqrt{\frac{\sum(x_i-\mu)^2}{N}}
    \end{equation}
    where $x_i$ is the values of the population, and $\mu$ the mean;
    \item \emph{coefficient of variation} is a measure of dispersion of a probability distribution. 
    The coefficient of variation ($CV$) is computed by the following equation:
    \begin{equation}
        CV = \frac{\sigma}{\mu}
    \end{equation}
    \item \emph{quartile coefficient of dispersion} is a statistics which measures dispersion. Since it is based on quantile information, it is less sensitive to outliers than other measures, such as the coefficient of variation.
    The quartile coefficient of dispersion ($QCD$) is computed by the following equation:
    \begin{equation}
        QCD = \frac{Q_3 - Q_1}{Q_3 + Q_1}
    \end{equation}
    where $Q_1$ and $Q_3$ are the value of the first quartile and the third quartile, respectively.
\end{enumerate}

\section{Execution Plan}
\label{sec:execution_plan}
In this section, we expose our approach to detect energy regression using developers' tests.

\subsection{Approach}
\label{subsec:execution_plan:approach}
Consider two versions of the same program $v_1$ and $v_2$, and code changes (commit), where $v_1$ and $v_2$ are the versions of the program before applying the code changes and the version of the program after applying the commit, respectively.
Our approach builds on 4 main steps:
1) it selects all the tests that cover the code changes,
2) it instruments the selected tests with specific probes to monitor key metrics for our study,
3) it executes each instrumented tests to collect the required measurements,
4) it computes the delta of the energy consumption introduced by the code changes.

The remainder of this section details the above key steps:
\autoref{subsubsec:execution_plan:approach:test-selection} details the test selection;
the test instrumentation is explained in \autoref{subsubsec:execution_plan:approach:test-instrumentation};
we describe the test execution in \autoref{subsubsec:execution_plan:approach:test-execution};
\autoref{subsubsec:execution_plan:approach:delta-computation} describes how the impact of code changes is approximated by the delta of the measurements of $v_1$ and $v_2$.
Then, \autoref{subsubsec:execution_plan:approach:analysis} exposes how we plan to perform the analysis of the results to decide if code changes should be tagged as ``breaking''.

In addition to this, \autoref{subsubsec:execution_plan:approach:fault_localization} details the technique we plan to adopt to locate the most suspect code changes that would be could cause the energy consumption drift.
Eventually, \autoref{subsubsec:execution_plan:approach:mutation} present how do we plan to perform the energy consumption mutation.

\subsubsection{Test selection}
\label{subsubsec:execution_plan:approach:test-selection}
The first step of our approach is to select the tests that will be used to detect energy regression.
To do this, we exploit three components: two versions of the same program and the code changes.
First, we compute the code coverage of the test suite for both versions.
At the end of this first step, we know the line coverage for each tests---\ie which test executes which line of the program.
Then, we select the tests according to the version and the operations of the code changes as follows:
we select from $v_1$ all the tests that executed the deletions of the code changes;
we select from $v_2$ all the tests that executed the additions of the code changes;
from this selection, we then discard all the modified tests, because the modification in tests might impact our measurements, \eg the software energy consumption, between both versions, while the goal is to measure the impact of the source code changes;
then, we take the union of both test subsets to be instrumented in the next step.

\subsubsection{Test instrumentation}
\label{subsubsec:execution_plan:approach:test-instrumentation}
The second step of our approach is to instrument the tests selected in the previous step to monitor measurements of interest.
This is can be done by injecting probes into the selected tests.
We inject two probes: one to start the monitoring and one to complete the monitoring by collecting appropriate metrics.
These start and stop will be automatically triggered when each test starts and completes (or fails), respectively.

\subsubsection{Test execution}
\label{subsubsec:execution_plan:approach:test-execution}
The third step of our approach is to execute the tests instrumented in the previous step.
Because of the uncertainty of energy consumption, which is a physical measure, the execution of the tests should be repeated to consider as accurate as possible measurements.
Furthermore, there is a lot of mechanisms interferes and cannot be, or are nearly impossible, to control.
These mechanism, such as runtime optimization, prediction, etc. can introduce noises in the collected measurements.

\subsubsection{Delta computation}
\label{subsubsec:execution_plan:approach:delta-computation}
The fourth step is to compute the delta for each variable of each test.
This is done to approximate the impact of code changes on the energy consumption of the program.
We first compute the test-wise deltas of each variable monitored during the test execution.
That is to say, we apply the formula:
\begin{equation}
    \forall t \in IS, \Delta_v(t) = V(t_2) - V(t_1)
\end{equation}
where $IS$ is the set of instrumented tests, $V(t_i)$ is the value of the monitored variable---\ie energy consumption, duration, number of instructions, etc. see \autoref{sec:variables} for more information, of the test $t$, for the version $i$---\ie $1$ or $2$.
$\Delta_V(t)$ is the delta of the considered variable $V$ of the test $t$ between the two versions of the program.

\subsubsection{Result analysis}
\label{subsubsec:execution_plan:approach:analysis}
The last step is to analyze the collected measurements associating the delta and the executed tests to eventually tag the associated code changes as ``breaking''.
In this section, we only consider the tests that have been selected by the approach and the lines that are included in the code changes---\ie lines that have been removed or added.
We consider $n$ tests, referred as $t_0, t_1, ... ,t_n$ and $m$ lines, referred as $l_0, l_1, ..., l_m$.
For the sake of clarity, we omit to precise which variables is used in these calculations---\eg duration, energy consumption, number of instructions---as these computations can be performed individually.

For each $c$ in $commits$, for each $v$ in $V$, for each $t$ in $IS$, we have $\Delta^c_v(t)$.
We recall that, if $\Delta^c_v(t) > 0$, it means that the code changes $c$ increases the energy consumption, which can be the energy consumption expressed in joules or approximated by either the time or performance counters, such as the number executed instructions, of the program when executing the test $t$.

As a test suite can include more tests increasing the SEC than tests decreasing it, we do not rely only on the computation of $\sum_{t=t_0}^{t_n}(\Delta^c_v(t))$.

For example, a code change modifies two lines of the program, $l_1$ and $l_2$, where $l_1$ increases the energy by $x\,\joule$ and $l_2$ decreases it by $y\,\joule$, with $y = 5 \times x$.
If there are 5 tests that (only) cover $l_1$ and 1 test that (only) covers $l_2$, we argue that it means that $l_1$ is more ''severely impacting`` than $l_2$ because $l_1$ has the largest ratio of executions among all the executed lines.
In other words, as $l_1$ is more covered than $l_2$, we consider that $l_1$ is more important.
Therefore, $L_1$ has more impact on the overall energy consumption of the program and this impact must be taken into account when the deciding if code changes are introducing a energy regression.
In this particular example, $\sum_{t=t_0}^{t_n}(\Delta^c_v(t)) = 0$, since $y = 5 \times x$ and there are 5 tests that cover (only) $l_1$ (consuming $x\,\joule$) and 1 test that (only) covers $l_2$ (consuming $y\,\joule$).
This means that taking roughly $\sum_{t=t_0}^{t_n}(\Delta^c_v(t))$ is not sufficient to decide if code changes are increasing or not the overall energy consumption.

We argue that the $\Delta(t)$ of each test $t$ should be weighted according to the proportion of the executions of lines over all the executions of all lines.
That is to say, for each line $l$, for each test $t$, we compute $exec(l,t)$ that gives the number of times the line $l$ is executed by the test $t$.
Then, for each line $l$, we compute:
\begin{equation}
    \theta(l)=\sum_{t=t_0}^{t_n} exec(l,t)
\end{equation}
\begin{equation}
    \Theta=\sum_{l=l_0}^{l_m}\theta(l)
\end{equation}
\begin{equation}
    \phi(l)=\theta(l)/\Theta
\end{equation}
where $\theta(l)$ is the total number of times the line $l$ is executed by the selected tests;
$\Theta$ is the total number of times all the lines are executed by the tests;
and $\phi(l)$ is the ratio of the total number of times the line $l$ is executed by the tests over the total number of times all the lines are executed by the tests.

Then, for each test $t$, we compute:
\begin{equation}
    \omega(t)=\sum_{l=l_0}^{l_m} \phi(l) \times exec(l,t)
\end{equation}
\begin{equation}
    \Omega(t)=\Delta(t) \times \omega(t)
\end{equation}
\begin{equation}
    \Delta_{\Omega}=\sum_{t=t_0}^{t_n} \Omega(t)
\end{equation}
where $\omega(t)$ is the weight that will be applied to the delta of the test $t$ computed from the sum of the $phi(l)$ for the line $l$ that the test $t$ executes;
$\Omega(t)$ is the delta of the test $t$ weighted by the weight of the test $t$; 
and $\Delta_{\Omega}$ is the delta of the code changes.

Let us reconsider the above example and apply the above equations results in the following:
$\theta(l_1)=5$, $\theta(l_2)=1$, and $\Theta=6$.
Then, $\phi(l_1)=0.84$ and $\phi(l_2)=0.17$;
$\sigma(t_1,t_2,t_3,t_4,t_5)=0.84$, and $\sigma(t_6)=0.17$;
$\omega(t_1,t_2,t_3,t_4,t_5)=0.84$, and $\omega(T_6)=-0.84$.
Eventually, $\Delta_{\Omega}=3.34$ as $\Delta_{\Omega} > 0$, then the code changes would be tagged as ``breaking'', while the rough $\sum_{t=t_0}^{t_n}(\Delta^c_v(t))$ would give $\Delta=0$.

\subsubsection{Most suspect code changes}
\label{subsubsec:execution_plan:approach:fault_localization}
When the code changes are considered as ``breaking'', we propose to rank the changed lines according to their suspiciousness.
This is done to hint developers on the potential cause of energy consumption drift.
To do this, we propose to rely on spectrum-based fault localization techniques from the state of the art~\cite{7390282}.
In particular, we propose to use techniques, such as \textsc{Tarantula}~\cite{Jones02visualizationof}, which ranks the software entities (methods, classes, lines, etc.) according to their number of executions by failing test cases.
In our context, the failing tests would be the ones that have their $\Delta_E(t)$ greater than $0$---\ie the test cases increasing the SEC.

\subsubsection{Energy consumption mutations}
\label{subsubsec:execution_plan:approach:mutation}
For executing our experimental protocol, we need apply code mutations that increase the energy consumption of a program.
This mutation consists in injecting an energy payload as an invocation of method \texttt{consumeEnergy}, described in \autoref{algo:consume_energy_mutant}, within each targeted method.

\begin{algorithm}
	\caption{Function to consume a given energy payload when mutating programs.}
	\label{algo:consume_energy_mutant}
	\begin{algorithmic}[1]
		\Require{Energy consumption probe: $P$}
		\Require{Energy payload to consume ($\mu\joule$): $E$}
		\Require{Seeded random generator: $R()$}
		\Require{Current time: $t()$}
		\State{$T \gets P.startMonitoring() + E$}
		\State{$random \leftarrow R(t())$}
		\While{$P.getCurrentEnergyConsumed() < T$}
		    \State{$random \leftarrow R(random)$}
		\EndWhile
		\State{$P.stopMonitoring()$}
	\end{algorithmic}
\end{algorithm}

The method \texttt{consumeEnergy} first starts the monitoring of the energy consumption (line~1) and estimate the threshold consumption $T$ by suming the energy payload $E$;
then, it initializes a variable called $random$ with a random value, using the seeded random generator with the current time as seed (line~2);
It keeps looping as long as the current energy consumed is less than the energy threshold (line~3);
at each iteration of the loop, it reassigns to the variable $random$ a new random value, using the seeded random generator with the current value of the $random$ variable as seed (line~4);
when the condition of the loop is met, the method \textsc{consumeEnergy} stops the monitoring and returns.

This method allows us to:
1) artificially increase the energy consumption of a program without altering its functional behavior;
2) configure the exact amount of energy to be consumed by the injected mutation;
3) bypass runtime optimizations as the value of the random number cannot be cannot be predicted;

\subsection{Implementation}
\label{subsec:execution_plan:implementation}
In our study, we focus on Java and Maven projects.
We restrict our study to this technology as we have a strong background in Java projects using Maven.
Also, we can rely on previous works that proposed relevant artifacts to help us.
While these choices reduce the time spent for our experimentation, we believe that our approach can be replicated to any technology and is not restricted to Java.

Our approach required various automated procedure.
For its implementation, we enhanced existing artefacts from the state of the art and implemented the missing parts.

\subsubsection{Test Selection}
For test selection, we enhanced the prototype \emph{diff-test-selection},\footnote{\url{https://github.com/STAMP-project/dspot/tree/master/dspot-diff-test-selection}} that has been developed by Danglot~\etal~\cite{danglot:hal-03121735} to select the tests that execute the code changes and amplify them.
Here, we make the artifact more accurate by implementing our test selection algorithm, as described in \autoref{subsubsec:execution_plan:approach:test-selection}, within \emph{diff-test-selection}.
In our experimentation, we decide to select the tests that executes specifically the code that have been changed in order to save computation time.
However, we plan to apply, for at least one project, the approach on all the test and on the subset of the selected tests.
This is will be done in order to assess the effectiveness of the test selection and therefore verify that selecting the unit tests to be considered according to the code changes is relevant.

\subsubsection{Test Instrumentation}
For test instrumentation, we first rely on the existing libraries \emph{JJoules}\footnote{\url{https://github.com/Mamadou59/j-joules}} and \emph{JUnit-JJoules}\footnote{\url{https://github.com/Mamadou59/junit-jjoules}} to collect metrics---\eg energy consumption, durations.
\emph{JJoules} uses RAPL~\cite{6337489} to collect measurements on the energy consumption of various physical components, such as CPU and DRAM, while \emph{JJoules} relies on the standard Linux library \emph{perf} to collect the various performance counters, such as the number of executed instructions, number of cycles, etc.
We implemented \emph{Diff-JJoules} as a set of Maven plugins to automatically instrument the tests of a project---\ie injecting \emph{JUnit-JJoules} into the tests, as described in \autoref{subsubsec:execution_plan:approach:test-instrumentation}.

\subsubsection{Test Execution}
For test execution, we relies on Maven, which provides all the features to repeat multiple executions of each test on both versions.

\subsubsection{Diff JJoules}
\emph{Diff-JJoules} also provides Python scripts to automate the whole approach on a specific use case.
\emph{Diff-JJoules} has been developed to be used in the CI workflow.
Thanks to the scripts, \emph{Diff-JJoules} is able to exploit the URL of a project---\eg a GitHub project---and two SHAs of commits, which are references to a specific version of the program, in order to apply our approach.
This is done as follows:
first, we clone twice the project, using the URL;
second, we set one of the cloned version of the project to $SHA(v_1)$, the other to $SHA(v_2)$;
third, \emph{Diff-JJoules} applies \emph{diff-test-selection} using $v1$ and $v2$ to select the test that execute the changes;
fourth, \emph{Diff-JJoules} injects JUnit-JJoules probes in the selected tests;
fifth, \emph{Diff-JJoules} uses maven to execute the tests;
sixth, \emph{Diff-JJoules} provides Python scripts to compute the energy deltas.

\subsubsection{Energy consumption mutations}
We implement our energy consumption mutations described in \autoref{subsubsec:execution_plan:approach:mutation}, as a Maven plugin relying on the library of Java code analysis and transformation Spoon~\cite{Spoon}.
This Maven plugin injects the method \texttt{consumeEnergy} and its invocation in the targeted methods.

\subsubsection{Analysis and Spectrum-based fault localization}
The two remaining parts to be implemented are the result analysis and its various computations, and the fault localization technique.
We plan to implement these as Maven artifacts to be easily integrated within the CI workflows.
We can use state-of-the-art implementation of \textsc{Tarantula}\footnote{\url{https://github.com/spideruci/Tarantula}} for this purpose.

\section{Datasets}
\label{sec:datasets}
To execute our experimental protocol, we need Java projects that use Maven as dependencies and routines manager, and we need code changes on these projects.
We decided to rely on an existing dataset built by Ourmani~\etal~\cite{ournani:hal-03202437}, as these projects are developed in Java, are using Maven and are providing a long development history: the commits of their Github repository.
These open-source projects are complex and large enough to generalize the results.
Our study also requires commits with specific properties applying to the source code, so we will not consider commits that modify the documentation or configuration files, for example.
In addition to this, we need to use only tests that cover the code changes, we are not interested in the tests that are unrelated to the code changes.
Eventually, we need tests that can be executed against the two versions of the program: before the commit and after applying the commit.
These requirements are the criteria that we will use to build our dataset of projects and versions (commits) of these projects.

\section{Threats to Validity}
In this section, we expose the validity threats that we identified so far.
A first threat relates to the delta consistency.
That is to say, the approach would bring contradictory line coverage and software energy consumption.
For example, only one line has been modified and it is executed by two tests the same number of times.
The deltas of each test are respectively $1000$ and $-1000$.
In this case, it would be impossible to decide if the code changes are considered as ``breaking'' by our approach.
However, we believe that this case will be exceptional and unlikely to happen in practice.

A second threat relates to the stability of the measurements.
It is well-known that energy measurements captured by power meters, like RAPL, are unstable.
To reduce the noise introduce by external events, we followed the guidelines proposed by Ourmani~\etal~\cite{ournani:hal-02892900}.
In addition to this threat, we cannot guarantee that our implementation and the integrated libraries, such as \emph{Diff-test-selection}, \emph{JJoules} or \emph{JUnit-JJoules} are without any bug.
However, these implementation are open-source and/or tested which might mitigate the risks.

As a third threat, our approach strongly depends on the quality of the test suites.
Therefore, we will consider open-source projects that report on at least 80\% of code coverage to mitigate this threat.

A fourth threat is that unit test may not be a strong representation of the production workload.
However, we place our approach in the context of continuous integration and code changes (commits).
The goal is to give hint to developers regarding the impact of the code changes on the energy consumption of the changes lines.
This is why we believe that weighting the energy consumption of tests by the weight of lines they execute is a reasonable approximation of the impact of the commits on the energy consumption.

A final threat is the worthfulness of the approach.
In the one hand, our approach might help saving some energy.
In the other hand, our approach consumes also energy.
In the case of the approach consumes more energy than it allows developers to save energy, the approach seems to waste resources instead of filling its purpose.
However, in the current state of the experimentation, we are not yet interested in observing the cost of our approach and leave as future works, as it would require a completely different setup, including usage profile of targeted projects.

\section{Related Work}
Pereira~\etal~\cite{PEREIRA2020110463} devised a technique called \emph{SPectrum-based Energy Leak Localization} (SPELL) that highlights energy hotspots in a program.
SPELL relies on fault localization techniques that collect software entities (methods, classes, lines, etc.) that are executed by use cases.
They implemented SPELL in Java, and experiment it on 5 projects to demonstrate that SPELL can identify energy hotspots.
The evaluation showed that SPELL helped the developers to improve the energy consumption.

Mancebo~\etal~\cite{DBLP:journals/sqj/ManceboCG21} analyzed the relation between maintainability and energy consumption of different versions of Redmine.
They computed the maintainability using SonarQube and measured the energy consumption using an \emph{Energy Efficiency Tester} (EET) appliance.
They concluded that the number of code lines and the energy consumed is correlated.

Maia~\etal~\cite{DBLP:conf/kbse/Maia0SP20} proposed the concept of energy debt, which is the amount of energy that a system consumes over time, due to energy code smells.
In this work, the authors defined a set of Android energy code smells, and developed \textsc{E-Debitum}, a SonarQube plugin that computes the energy debt between versions of Android applications.
They evaluated their approach on 3 Android applications, and proved that it can be applied on real-world applications by demonstrating the evolution of their energy debt over time.

Luo~\etal~\cite{10.1145/2901739.2901765} proposed a technique, called \textsc{PerfImpact}, which recommend inputs and code changes related to performance regressions.
To do this, they use the combination of search-based and change impact analysis.
They evaluated \textsc{PerfImpact} on 2 open-source projects.
They showed that \textsc{PerfImpact} is able to identify performance issues between two versions of a program.
\textsc{PerfImpact} is focused on determining the specific inputs and the code changes that trigger a performance regression, in term of time execution, while in our studies we plan to use the input provided by developer tests to observe if there are energy regression.
Both of our approaches are taking place in the context of continuous integration and performance regression, even if our approach is narrowed to energy consumption regression, which can be seen as special case of performance regression.

Chen~\etal~\cite{8094434} perform an exploratory study on the code changes that introduce performance regressions.
They perform an evaluation on $1126$ commits of Hadoop and $135$ commits of RxJava.
The evaluation reported on $243$ and $91$ commits introducing performance regression in Hadoop and RxJava, respectively.
They also identified 6 root-causes of performance regressions introduced by code changes.

Ding~\etal~\cite{ding2020towards} study the use of tests in the release pipeline as performance tests of Hadoop and Cassandra, two open-source projects. 
They show that, for $102$ out of $127$ of the performance issues, there is at least one test that can be used to spot a performance improvement.
This study comforts our hypothesis that developers' tests can be used to detect energy regression, as energy consumption is correlated (not only to) to the performance of the program.

Hindle \etal~\cite{10.1145/2597073.2597097} devise GreenMiner.
GreenMiner is a dedicated hardware/software mining software repositories test harness.
GreenMiner has been created to study relationships between code changes and power consumption.
The main difference is that GreenMiner is using physical measurements and works only for Android applications, while we use RAPL to measure the energy consumption and the approach could be applied on any software.

Hindle~\cite{hindle2015green} presents Green mining.
Green mining attempts to estimate empirically the impact of software change on energy consumption in order to guide developers to reduce the energy consumption of the Software.
Green mining comes also with an abstract methodology that gives key steps in order to setup the approach on an application.
The author performs an evaluation on multiple branches of Firefox, on release history of Azureus/Vuze and a study the effect of multiple versions a the library libTorrent on power consumption of its client application rTorrent.
The key difference between Green mining and our approach is that we try to identify the impact on energy consumption of code changes, while Green mining is working at a more coarse-grained level.

Chowdhury \etal~\cite{10.1145/2901739.2901763} propose GreenOracle, a model to predict the energy consumptions.
This model is trained using a large corpus of Android applications, their system calls dynamic traces, CPU utilization and GreenMiner.
They collected $984$ versions of $24$ different Android applications and showed estimation of energy consumption with less than 10\% error.

Romansky~\etal~\cite{8094428} propose an approach, called Deep Green, to construct models that use software performance measurements to predict instantaneous energy consumption based on time series.
The authors uses GreenMiner to train models to estimate software energy consumption.

The main difference between our approach and the two last works, GreenOracle and Deep Green, is that we measure the energy consumption using developer tests, while GreenOracle is based on predictions of the energy consumption of program versions.

\bibliographystyle{IEEEtran}
\bibliography{IEEEabrv,references}

\end{document}